\documentclass{emulateapj}
\slugcomment{Accepted for ApJ Letters }

\shorttitle{OGLE-2009-BLG-076S -- THE MOST METAL-POOR DWARF STAR IN 
THE GALACTIC BULGE}
\shortauthors{BENSBY ET AL.}

\newcommand\teff{T_{\rm eff}}
\newcommand\kms{\rm\,km\,s^{-1}}

\begin{document}

\title{
OGLE-2009-BLG-076S --- THE MOST METAL-POOR DWARF STAR IN THE 
GALACTIC BULGE\altaffilmark{1}
}
\author{
T. Bensby,\altaffilmark{2}
S. Feltzing,\altaffilmark{3}
J.A. Johnson,\altaffilmark{4}
A. Gal-Yam,\altaffilmark{5}
A. Udalski,\altaffilmark{6}
A. Gould,\altaffilmark{4}\\
C. Han,\altaffilmark{7}
D. Ad\'en,\altaffilmark{3}
and
J. Simmerer\altaffilmark{3}
}

\altaffiltext{1}{Based on observations made with the European Southern 
Observatory telescopes, Program ID 082.B-0453(B).}
\altaffiltext{2}{European Southern Observatory, Alonso de Cordova 3107, 
Vitacura, Casilla 19001, Santiago 19, Chile; {\tt tbensby@eso.org}}
\altaffiltext{3}{Lund Observatory, Box 43, SE-221\,00 Lund, Sweden;
{\tt sofia, daniela, jennifer@astro.lu.se}}
\altaffiltext{4}{Department of Astronomy, Ohio State University, 
Columbus, OH 43210, USA; 
{\tt jaj,\,gould@astronomy.ohio-state.edu}}
\altaffiltext{5}{Benoziyo Center for Astrophyics, Weizmann Institute of 
Science, 76100 Rehovot, Israel; {\tt galyam@wisemail.weizmann.ac.il}}
\altaffiltext{6}{Warsaw University Observatory, Warszawa, Poland; 
{\tt udalski@astrouw.edu.pl}}
\altaffiltext{7}{Department of Physics, Chungbuk 
National University, Chongju, Republic of Korea; {\tt cheongho@astroph.chungbuk.ac.kr}}

\begin{abstract}
Measurements based on a large number of red giant stars suggest a 
broad metallicity distribution function (MDF) for the Galactic bulge, 
centered on $\rm [Fe/H]\approx -0.1$. However, recently, a new 
opportunity emerged to utilize temporary flux amplification (by 
factors of $\sim100$ or more) of faint dwarf stars in the Bulge that 
are gravitationally lensed, making them observable with high-resolution 
spectrographs during a short observational window. Surprisingly, of 
the first 6 stars measured, 5 have $\rm [Fe/H]>+0.30$, suggesting a 
highly skewed MDF, inconsistent with  
observations of giant stars. Here we present a detailed elemental 
abundance analysis of OGLE-2009-BLG-076S, based on a high-resolution 
spectrum obtained with the UVES spectrograph at the ESO Very Large 
Telescope. Our results indicate it is the most metal-poor dwarf star 
in the Bulge yet observed, with $\rm [Fe/H]=-0.76$. Our results argue 
against a strong selection effect disfavoring metal-poor microlensed 
stars. It is possible that small number statistics is 
responsible for the giant/dwarf Bulge MDF discrepancy. Should this 
discrepancy survive when larger numbers of Bulge dwarf stars (soon to 
be available) are analyzed, it may require modification of our 
understanding of either Bulge formation models, or the behavior of 
metal-rich giant stars.  
\end{abstract}


\section{Introduction}

The metallicity distribution function (MDF) of the Galactic bulge
(hereafter, the Bulge) as measured by red giant stars
has a long controversial history. In the 1980s and 1990s studies of
red giant stars in Baade's window using low-dispersion spectroscopy, 
showed that the Bulge MDF was quite metal-rich at 
$\rm [Fe/H]\approx +0.2$ 
\citep[e.g.,][]{whitford1983,rich1988,terndrup1990}. 
Later, the mean metallicity was revised downward to $\rm [Fe/H=-0.25$
by the first high-resolution study of K giant stars in Baade's window
by \cite{mcwilliam1994}. Since then, results based on high-resolution 
spectroscopic studies 
of several hundred giant stars in or around Baade's window, show 
that the Bulge MDF peaks slightly below solar metallicity at 
$\rm [Fe/H]\approx-0.1$
\citep{fulbright2006,fulbright2007,cunha2006,cunha2007,cunha2008,
rich2005,rich2007,lecureur2007,zoccali2003,zoccali2008}. 
However, the recent progress in observing low-luminosity dwarf 
stars in the Bulge while they are being optically magnified during 
gravitational microlensing events appears to give markedly different 
results and 
have re-ignited the debate over the Bulge MDF. So far, detailed 
elemental abundances based on high-resolution spectroscopy for a 
total of six microlensed dwarf and subgiant stars have been published
\citep{johnson2007,johnson2008,cohen2008,cohen2009,bensby2009}.
Five of these six stars have high super-solar metallicities in 
the range $\rm +0.25<[Fe/H]<+0.55$ and it is only OGLE-2008-BLG-209S
that is somewhat metal-poor at $\rm [Fe/H]=-0.32$ \citep{bensby2009}. 
The average metallicity from the six microlensed Bulge stars is 
$\rm [Fe/H]\approx+0.29$, which is 0.4\,dex higher 
than the MDF obtained from giant stars.

OGLE-2008-BLG-209S is not just the only one of the six microlensed 
stars that turned out to be metal-poor but also the only one that 
turned out to be slightly evolved, with stellar parameters placing it 
on the subgiant branch. This led \cite{cohen2009} to speculate that 
OGLE-2008-BLG-209S is an exception to the rule, and that it 
should perhaps be excluded from the microlensed stellar sample. 
The average 
metallicity would then rise to $\rm [Fe/H]=+0.41$, i.e., even more 
discrepant when comparing to the giant star MDF.

As microlensing events have no preference of picking source stars
with a particular metallicity, observing enough events should provide
an unbiased estimate of the Bulge MDF. So far, the microlensing events
that have been observed in the Bulge lie at approximately the same
angular distance from the Galactic center as the red giant stars in
Baade's window. Hence, 
there is no reason to expect that their MDFs should differ. 
By randomly picking 6 stars, 40\,000 times, 
from the \cite{zoccali2008} sample of giant stars, \cite{cohen2009} 
found that in only 0.4\,\% of the cases, the randomly 
drawn red giant sample 
was as metal-rich as the sample of microlensed dwarf stars. The 
question was then which of the 
two tracers that give the correct picture: giant stars or the 
microlensed dwarf stars? 

In this Letter we present first results from a detailed abundance 
analysis of OGLE-2009-BLG-076S, the source star of another microlensing
event toward the 
Bulge. We focus here on the metallicity, the $\alpha$-elements Mg, Si,
and Ti, the stellar age, distance, and the star's kinematic properties. 
The full analysis and results for other $\alpha$-elements, light 
elements, iron-peak elements, $r$-, and $s$-process elements, will be 
presented in an upcoming paper.

\section{Observations and data reduction}

\begin{figure}
\resizebox{\hsize}{!}{
\includegraphics[bb=20 50 530 515,clip]{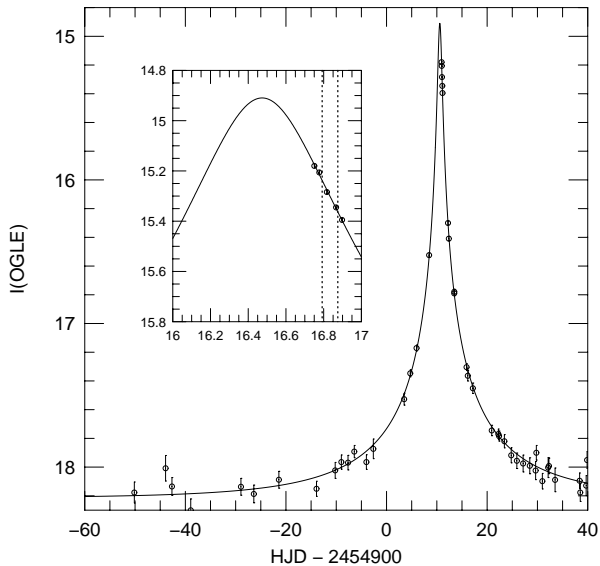}
}
\caption{OGLE photometry of the microlensing event. The vertical dotted
lines indicate the interval of the observations with VLT UVES.
\label{fig:events}
}
\end{figure}

On 2009 March 21, the OGLE early warning 
system\footnote{\tt http://ogle.astrouw.edu.pl/ogle3/ews/ews.html} 
\citep{udalski2003} identified OGLE-2009-BLG-076S as a possible 
high-magnification microlensing event toward the Bulge at
$(l,\,b)=(1.2,\,-2.5)$\,deg. Since the intrinsic source flux (inferred 
from the microlensing model) indicated that the star was a
dwarf star we triggered our ToO observations, without knowing the color 
of the star,
with the ESO Very Large Telescope 
(VLT) on Paranal. Due to the limited visibility of the Bulge in 
March, the target had to be observed toward the end of the night. 
Hence, OGLE-2009-BLG-076S was observed during the early morning hours 
beginning March 26 (MJD: 4916.29099), a few hours after 
peak brightness
(see Fig.~\ref{fig:events}). At maximum the light from the
source star was amplified by a factor of 68. 
Using the UVES spectrograph 
\citep{dekker2000short}, located at VLT UT2, configured with dichroic 
number 2, the target
was observed for a total of two hours, split into four 30 minute 
exposures. The resulting spectrum was recorded on three CCDs with 
wavelength coverages between 3760-4980\,{\AA} (blue CCD), 
5680-7500\,{\AA} (lower red CCD), and 7660-9460\,{\AA} (upper red CCD). 
A $1''$ slitwidth  yielded a spectral resolution  of $R\approx45\,000$.

\begin{figure}
\resizebox{\hsize}{!}{
\includegraphics[angle=270,bb=40 28 560 780,clip]{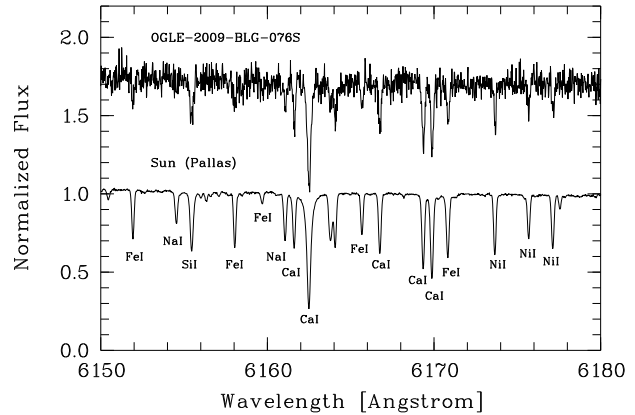}}
\caption{Normalized spectra of OGLE-2009-BLG-076S and the Sun 
(reflected from the asteroid Pallas), obtained with VLT UVES. 
The spectra have been shifted vertically and horizontally.
        \label{fig:spectrum}}
\end{figure}

The data were reduced with the most 
recent version of the UVES pipeline. The typical signal-to-noise ($S/N$)
ratio per pixel at 6000\,{\AA} in the lower red CCD is 
$\sim30$ (compare Fig.~\ref{fig:spectrum}). In the 
spectrum from the blue CCD, the $S/N$ was, however, too low to allow  
either secure identification of spectral line features or
accurate measurement of equivalent widths. Hence, the spectral 
region blue-ward of 
5680\,{\AA} was not used in the analysis.

Before the observation of the main target, 
we observed HR\,6141, a rapidly rotating B2V star, at 
an airmass similar to what was expected for the Bulge star, 
to divide out telluric lines in the spectrum. We also obtained 
a solar spectrum,
by observing the asteroid Pallas, that was used to normalize the
elemental abundances in OGLE-2009-BLG-076S to those in the Sun.
The $S/N$ per pixel at 6000\,{\AA} in the solar spectrum and the 
B star spectrum is $\sim 350$ and $\sim 500$, respectively.

Examples of the Bulge star spectrum and the solar spectrum
are shown in Fig.~\ref{fig:spectrum}.

\section{Analysis}

\subsection{Stellar parameters from spectroscopy}
\label{sec:analysis}

The determination of stellar parameters and calculation of elemental
abundances were done in the very same way as in Method 1 of 
\cite{bensby2009}, wherein 
full details of the analysis can be found, which is based on equivalent 
width measurements 
and one-dimensional LTE model stellar atmospheres calculated with 
the Uppsala MARCS code 
\citep{gustafsson1975,edvardsson1993,asplund1997}. 
The linelist is an expanded version of the linelist of 
\cite{bensby2003,bensby2005} and is fully given in 
Bensby et al.~(2010, in prep.). The effective temperature ($\teff$) is 
determined by requiring excitation balance of abundances from 
Fe\,{\sc i} lines, the microturbulence parameter ($\xi_{\rm t}$) by 
requiring balance of abundances from Fe\,{\sc i} lines 
versus the reduced equivalent width ($\log (EW/\lambda)$), 
and the surface gravity ($\log g$) from ionization balance between
abundances from Fe\,{\sc i} and Fe\,{\sc ii} lines. 

Due to shorter wavelength coverage and lower $S/N$ only
a fraction of the $\sim 250$ Fe\,{\sc i} and $\sim 30$ Fe\,{\sc ii} 
lines available in the above mentioned linelist could be measured
and utilized in the analysis. Hence, based on 57 Fe\,{\sc i} and 
7 Fe\,{\sc ii} lines we find that 
OGLE-2009-BLG-076S has $\teff=5877$\,K, $\log g = 4.30$, 
$\xi_{\rm t}=1.61$\,$\kms$, and an absolute Fe abundance of 
$\rm \log \epsilon (Fe) = 6.82$, or $\rm [Fe/H]=-0.76$.

\subsection{Uncertainties}
\label{sec:errors}

The errors in the stellar parameters are of standard nature, and we 
estimate them to be of the same magnitude as in 
\cite{bensby2009}, i.e., $\sigma\teff=200$\,K, 
and $\sigma\log g=0.2$. These errors will introduce a random error 
of 0.12\,dex in [Fe/H] \citep[see Table~6 in][]{bensby2009}. Further, 
the standard deviation around the mean Fe\,{\sc i} abundance is 
0.10\,dex. The corresponding formal error ($\sigma/\sqrt{N}$)
in the mean Fe abundance is 0.013\,dex, which is negligible
in comparison to the effects that the uncertainties in the stellar
parameters have on [Fe/H]. Hence, we estimate that the total error in 
[Fe/H] is 0.12\,dex.

\subsection{Microlensing techniques}

Using microlensing techniques \citep{yoo2004short}, based on $V$ and 
$I$ data taken with the SMARTS 1.3\,m telescope at CTIO, 
OGLE-2009-BLG-076S is estimated to have an intrinsic color of 
$(V-I)_{0}=0.69\pm0.05$, i.e., similar to the Sun, and an 
absolute magnitude 
of $M_{\rm V}=4.76-5\log(d_{\rm S}/d_{\rm clump})\pm0.07$. 
The last term originates from the fact that the microlensing technique 
assumes that the source star (S) lies at the same distance as the
clump stars, in which case this term would vanish.
The 
color-$\teff$ calibration by \cite{ramirez2005} then implies that the
effective temperature should be 5670\,K. This is 200\,K lower than the
 ``spectroscopic'' temperature that we derive. However, if the 
true temperature of OGLE-2009-BLG-076S is lower, the abundances 
derived from the Fe\,{\sc i} lines will also be lower, making the 
star even more metal-poor.

\section{Properties of OGLE-2009-BLG-076S}

\begin{figure}
\resizebox{\hsize}{!}{
\includegraphics[angle=-90,bb=283 40 560 455,clip]{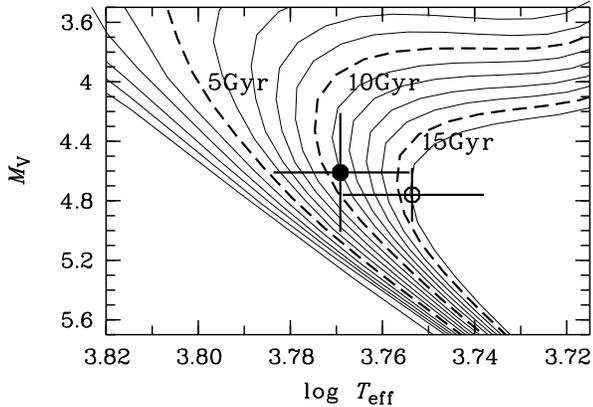}}
\caption{
         The Y$^2$ isochrones that were used to estimate the 
         stellar ages (see Sect.~\ref{sec:status}). Isochrones are 
         plotted in 1\,Gyr steps. 
         Filled circle represents the spectroscopic values (5870\,K and
         $M_{\rm V}=4.61$), while the open circle represents the values 
         from microlensing techniques. Horizontal error bars represent 
         200\,K in $\teff$ in both cases. In the microlensing case the
         error bar of 0.18\,dex in 
         $M_{\rm V}$ is taking account
         of the mean and dispersion of distance of microlens sources
         relative to the peak of the clump. 
         The spectroscopic uncertainty
         in $M_{\rm V}$ is estimated to be 0.4\,dex. 
         \label{fig:getage}}
\end{figure}

\subsection{Evolutionary status}
\label{sec:status}

The values of the effective temperature and surface gravity 
demonstrates that OGLE-2009-BLG-076S is a true dwarf star. 
Utilizing the fundamental relationship between surface gravity,
effective temperature, bolometric magnitude (see, e.g., Eq.~4 in 
\citealt{bensby2003}), we find an absolute magnitude of 
$M_{\rm V}=4.61$, in reasonable agreement with the value
based on microlensing techniques. 
Plotting the star on top of the Yonsei-Yale 
isochrones \citep{yi2001,kim2002,demarque2004}  we 
find an age of $11\pm4$\,Gyr,  for OGLE-2009-BLG-076S,
(see Fig.~\ref{fig:getage}). This old age is consistent with 
Bulge stars being an old population \citep[e.g.,][]{feltzing2000b}, 
and is similar
to what is found  for thick disk stars at the same metallicity 
\citep[e.g.][]{bensby2005,bensby2007letter2}. 
From the evolutionary tracks
of \cite{yi2003} we find that OGLE-2009-BLG-076S has a
mass of $\sim0.9$\,M$_{\odot}$.

\subsection{Radial velocity}

From the spectrum we measure a heliocentric radial velocity of 
$+128.7\,\kms$ for OGLE-2009-BLG-076S, which should be compared with 
the observed
mean and standard deviation of Bulge stars toward this direction,
$+27\pm 110\,\kms$ \citep{zhao1996,howard2008}.
Hence, OGLE-2009-BLG-076S is consistent with Bulge kinematics.

\subsection{Bulge membership}

Microlensed sources are not random Bulge sources, they
are heavily biased to come from more distant parts of the Bulge.
This is because the lens \underline{must} be in front of the source, 
and the probability of lensing is roughly proportional to the
square-root of the lens-source distance.  Most lenses are themselves
in the Bulge, particularly toward this direction, where the integrated
density of Bulge sources is extremely high.

Furthermore, the event is at $(l,b) = (1.2,-2.5)$\,deg, i.e.,
350\,pc below the Galactic center.  In this direction, the Bulge
(which is flattened by a factor of $\sim0.6$) has an effective width
of 600\,pc. Beyond this range, its density falls off very rapidly
\citep[see, e.g., the model by][]{han1995,han2003}.
Hence, a random source would have a very low probability of lying
in front of the red clump centroid. 

Given the low probability of the source star being a foreground 
disk star,
that the measured radial velocity of OGLE-2009-BLG-076S
is consistent with the bulk of Bulge stars, that it has an old age
consistent with the Bulge being an old stellar population, it is very 
likely that OGLE-2009-BLG-076S is a Bulge dwarf star.

\subsection{Elemental abundance ratios}
\label{sec:abundances}

\begin{figure}
\resizebox{\hsize}{!}{
\includegraphics[bb=60 160 590 510,clip]{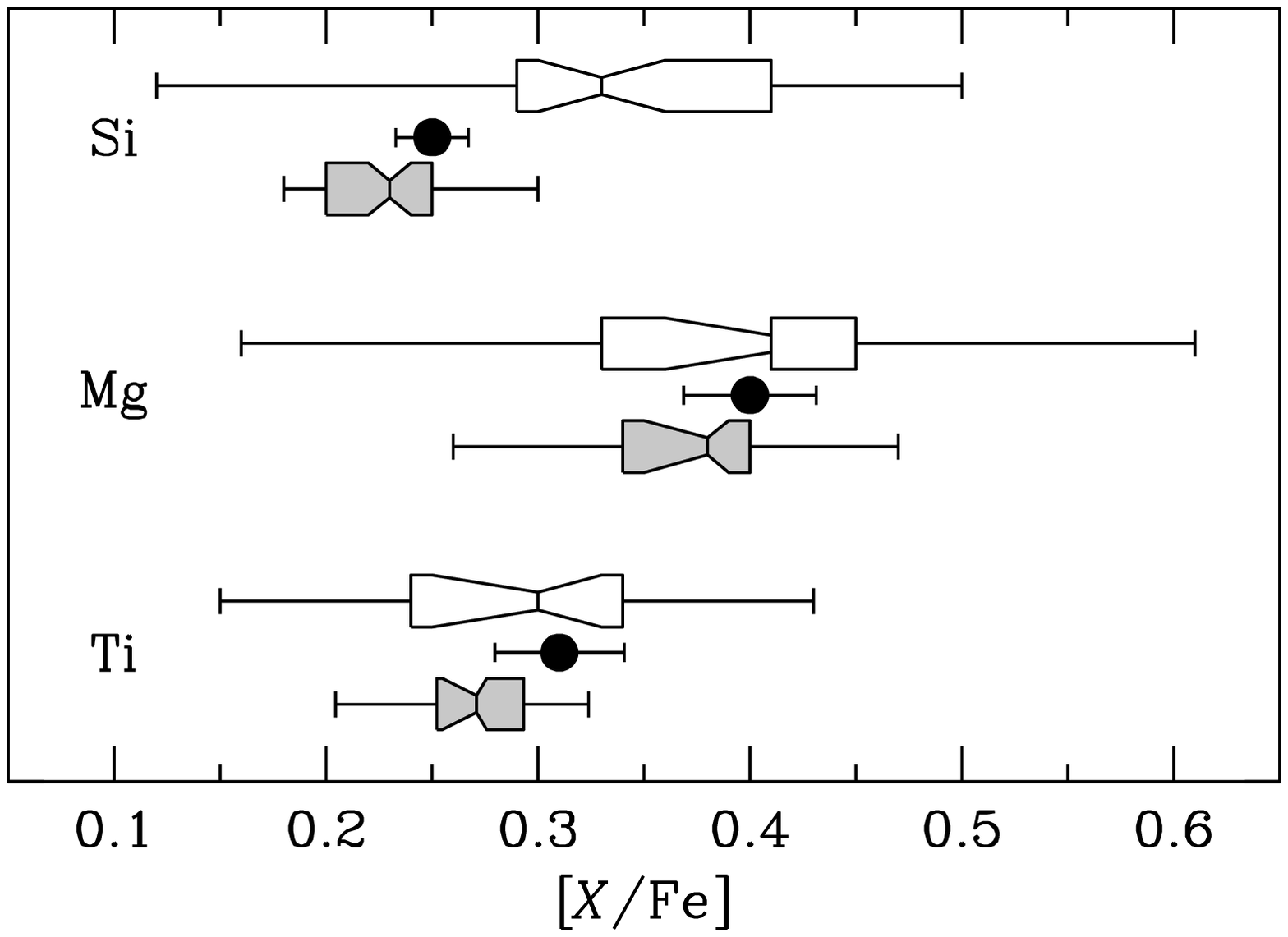}}
\caption{
        Boxplots showing the [$X$/Fe] distribution, for three 
        elements as indicated in the figure, for Bulge giant stars 
\citep[white boxes;][]{rich2005,rich2007,fulbright2007,lecureur2007}, 
        and thick disk dwarf stars 
\citep[grey boxes;][and Bensby et al.~2009, in prep.]{bensby2003,bensby2005,bensby2007letter2}.
        OGLE-2009-BLG-076S is marked by the filled circles, 
        where the error
        bars represent the formal error ($\sigma/\sqrt{N}$).
        In the boxplots the central vertical line represents the
        median value. The lower and upper quartiles are represented 
        by the
        outer edges of the boxes.
        The notches indicate the 95\,\% confidence intervals for the
        median value. The whiskers extend to the farthest data 
        point that
        lies within 1.5 times the inter-quartile distance.   
   \label{fig:halter}}
\end{figure}

The abundances of OGLE-2009-BLG-076S were normalized to those
of the Sun (as determined from the Pallas spectrum) on a line-by-line 
basis and then averaged for each element. The absolute iron abundance
measured in Sect.~\ref{sec:analysis} is then equal to 
$\rm [Fe/H]=-0.76$.

Figure~\ref{fig:halter} shows three $\rm [\alpha/Fe]$
abundance ratios in OGLE-2009-BLG-076S compared to the average 
abundance ratios for thick disk stars in the 
metallicity range $\rm -0.86<[Fe/H]<-0.66$ and to Bulge giant 
stars in the 
metallicity range $\rm -1<[Fe/H]<-0.3$. 
Compared to the Bulge giant stars, OGLE-2009-BLG-076S shows the same
degree of Mg and Ti enhancements, while it has a lower enhancement 
in Si.
Compared to the thick disk dwarf stars, OGLE-2009-BLG-076S has similar
enhancements in all three elements, with a hint of being
slightly more enhanced than the median thick disk. 
These $\alpha$-to-iron over-abundances at $\rm [Fe/H]=-0.76$
indicate a period of intense star formation in the Bulge where
chemical enrichment mainly is due to massive stars, exploding as
core-collapse supernovae. Furthermore, the solar-type abundance ratios 
$\rm [\alpha/Fe]\approx 0$ for the other five microlensed dwarf stars 
at super-solar metallicities 
\citep{bensby2009,johnson2007,johnson2008,cohen2008,cohen2009} 
is consistent with the drop in $\alpha$-enhancement above solar
metallicity observed in giant stars 
\citep[compare Fig.~10 in][]{bensby2009}.

\section{Discussion and conclusion}

OGLE-2009-BLG-076S is the currently most metal-poor dwarf star 
in the Bulge for which a detailed elemental abundance 
analysis based on high-resolution spectroscopy has been performed. 
It has a metallicity
of $\rm [Fe/H]=-0.76$, $\rm [\alpha/Fe]$ abundance ratios consistent
with in situ Bulge giant stars, and an old age of 11.5\,Gyr.
This result demonstrates that the Bulge does indeed contain metal-poor
dwarf stars, as predicted by the presence of metal-poor giant stars, 
and that it is possible to observe them while they
are being gravitationally microlensed. 

\begin{figure}
\resizebox{\hsize}{!}{
\includegraphics[bb=20 160 590 470,clip]{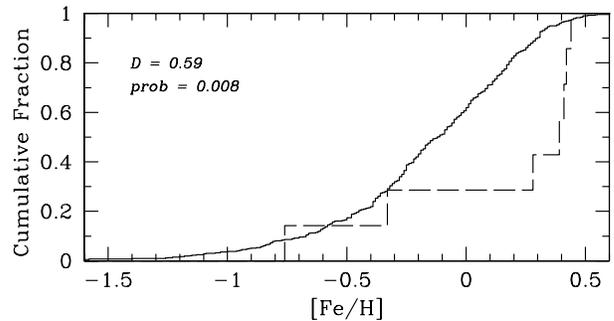}}
\caption{Two-sample Kolmogorov-Smirnov test between the MDF from 
the \cite{zoccali2008} giant stars (full line) and the MDF from 
the seven microlensed dwarf and subgiant stars (dashed line).  
Indicated in the figure is the $D$ statistic (maximum distance
between the two distributions) and the corresponding
probability that the two MDFs have been drawn from the 
same distribution.
         \label{fig:kstest}}
\end{figure}

The now seven microlensed dwarf and subgiant stars in the Bulge
have an average metallicity of $\rm \langle [Fe/H]\rangle = +0.12$
(using [Fe/H] values from \citealt{bensby2009} for the first four 
events).
A two-sided Kolmogorov-Smirnow test with the giant stars from
\cite{zoccali2008} gives a very low probability that the
MDFs from the dwarfs and the giants are the same 
(see Fig.~\ref{fig:kstest}).

There are essentially three possibilities to explain the giant/dwarf 
Bulge MDF discrepancy: {\bf (i)}  Due to systematic errors in either 
giant or dwarf metallicity measurements, the identical underlying MDF 
appears different. \cite{cohen2009} discuss this, including the size 
of the required errors, and find it very unlikely that the 
analysis methods 
are faulty; {\bf (ii)} The dwarf and the giant MDFs \underline{are} 
different and have been correctly measured to be different. The 
discrepancy is caused by mass loss \citep[as discussed by][]{cohen2009}, 
or because of different spatial distributions and chemical
inhomogeneities 
in the Bulge; {\bf (iii)} The dwarf and giant MDFs are identical, but 
small number statistics for the dwarf stars happen to cause them look 
different; {\bf (iv)} or some other unknown mechanism.

The new results for OGLE-2009-BLG-076S favor a combination of
(ii) and (iii). While the giant star MDF is well-defined, it could 
also be,
as suggested by \cite{cohen2008} that the most metal-rich giant stars
lose their entire envelope before reaching the RGB tip. Hence the
most metal-rich giants are missing, and the MDF is somewhat shifted
to lower [Fe/H]. This in combination with the small sample,
perhaps too small to be representative of the Bulge MDF,
of now seven microlensed stars: five very metal-rich dwarf stars;
one subgiant star just below solar metallicity; and one metal-poor
dwarf star, makes the difference between the giant star MDF and the MDF
from the sample of microlensed stars too large. 
However, with  OGLE-2009-BLG-076S, we have shown that
metal-poor dwarf stars exist in the Bulge, and that, if enough 
microlensing
events are observed, the differences between the giant and dwarf MDFs 
could in principle diminish to a level where the differences only 
are due to
the effects of mass-loss of the most metal-rich giant stars. 

\acknowledgements

 S.F. is a Royal Swedish Academy of Sciences Research Fellow supported 
 by a grant from the Knut and Alice Wallenberg Foundation. Work by A.G. 
 was supported by NSF Grant AST-0757888.  J.S. is supported by a Marie 
 Curie Incoming International Fellowship.
 A.U. acknowledges  support by the Polish MNiSW grant N20303032/4275. 
 A. G.-Y. acknowledges support by the Israeli Science Foundation, an
 EU Seventh Framework Program Marie Curie IRG fellowship and the
 Benoziyo Center for Astrophysics,
 a research grant from the Peter and Patricia Gruber Awards,
 and the William Z. and Eda Bess Novick New Scientists Fund at the
 Weizmann Institute. C.H. was supported by the Creative Initiative 
 program (2009-008561) of
 Korean Science and Enginerring Foundation.


\end{document}